\documentclass[conference]{IEEEtran}
\IEEEoverridecommandlockouts
\usepackage{xcolor}
\usepackage{amsmath}
\usepackage{dsfont} % Double-stroke indicator with an embedded vector font.

\usepackage{tikz}
\usepackage{pgfplots}
\usepackage{pgfplotstable}
\usepgfplotslibrary{fillbetween}
\usepgfplotslibrary{groupplots}
\pgfplotsset{compat=1.17}
\usetikzlibrary{arrows.meta}
\usepackage[font=footnotesize]{caption}
\usepackage{lipsum}

\usepackage[range-phrase=--,per-mode=symbol-or-fraction,range-units=single,list-units=single,detect-all]{siunitx}
\DeclareSIUnit\pp{pp}
\DeclareSIUnit\percentagepoint{percentage\ points}
\usepackage{csquotes}
\usepackage[backend=biber,style=ieee,doi=false,isbn=false,mincitenames=1,maxcitenames=2]{biblatex}
\DeclareFieldFormat{sentencecase}{#1} % never apply sentence casing, even if bibtex field is unprotected
\DeclareFieldFormat{titlecase}{#1} % never apply title casing, even if bibtex field is unprotected
\usepackage{xpatch}
\xpatchbibmacro{textcite}{\addspace}{\addnbspace}{}{}
\xpatchbibmacro{Textcite}{\addspace}{\addnbspace}{}{}

\DefineBibliographyStrings{english}{
	andothers = et~al\adddot\addspace
}

\usepackage[hidelinks]{hyperref}
\usepackage[capitalise,noabbrev]{cleveref}
\Crefformat{figure}{#2Fig.~#1#3}
\Crefformat{equation}{#2Eq.~(#1)#3}

\ifCLASSINFOpdf
\else
\fi

\usepackage{graphicx}

\usepackage{acro}
\usepackage{mfirstuc}
\DeclareAcronym{MC}{short=MC, long=molecular communication}
\DeclareAcronym{FSI}{short=FSI, long=Fluid-Solid Interaction}
\DeclareAcronym{PS}{short=PS, long=Peak Systole}
\DeclareAcronym{ED}{short=ED, long=Early Diastole}
\DeclareAcronym{LD}{short=LD, long=Late Diastole}
\DeclareAcronym{CIR}{short=CIR, long=Channel Impulse Response}
\DeclareAcronym{TX}{short=TX, long=transmitter}
\DeclareAcronym{RX}{short=RX, long=receiver}
\DeclareAcronym{LES}{short=LES, long=Large Eddy Simulation}
\DeclareAcronym{DNS}{short=DNS, long=Direct Numerical Simulation}
\DeclareAcronym{RAS}{short=RAS, long=Reynolds Averaged Simulation}
\DeclareAcronym{3D}{short=3D, long=Three-Dimensional}
\DeclareAcronym{IoBNT}{short=IoBNT, long=Internet of Bio-Nano Things}
\DeclareAcronym{SPION}{short=SPION, long=Superparamagnetic Iron-Oxide Nanoparticle}
\DeclareAcronym{MPPICFoam}{short=MPPIC, long=Multi-Phase Particle-In-Cell Foam}
\DeclareAcronym{CVS}{short=CVS, long=Cardiovascular System}
\DeclareAcronym{BNM}{short=BNM, long=Bio-Nano Machine}
\DeclareAcronym{SOI}{short=SOI, long=Start of Injection}
\DeclareAcronym{CFD}{short=CFD, long=Computational Fluid Dynamics}
\DeclareAcronym{ICA}{short=ICA, long=internal carotid artery}
\DeclareAcronym{ECA}{short=ECA, long=external carotid artery}
\DeclareAcronym{CAD}{short=CAD, long=computer-aided design}
\DeclareAcronym{KNN}{short=KNN, long={$k$-nearest-neighbor}}
\DeclareAcronym{MLP}{short=MLP, long=multilayer perceptron}
\DeclareAcronym{MAE}{short=MAE, long=mean absolute error}

\usepackage{titlesec}
\usepackage[absolute]{textpos}
  
\TPMargin{5pt}

\newcommand{\copyrightstatement}{
    \begin{textblock}{0.84}(0.08,0.01)    % tweak here: {box width}(leftposition, rightposition)
         \noindent
         \footnotesize
         Personal use of this material is permitted. This work has been submitted to the IEEE for possible publication. Copyright may be transferred without notice, after which this version may no longer be accessible.
    \end{textblock}
}

\begin{document}

\copyrightstatement

\title{Molecular Communication for Plaque Sensing in a Vascular Bifurcation with Receiver Constraints\
\thanks{\textsuperscript{*}The authors contributed equally to this work.\\
A. Wietfeld and W. Kellerer are with the Chair of Communication Networks, Technical University of Munich, Germany (e-mail: \{alexander.wietfeld,wolfgang.kellerer\}@tum.de).\\
L. Wille and J. Kirchner are with the Institute for Smart Electronics and Systems, Friedrich-Alexander-Universität Erlangen-Nürnberg, Germany (e-mail: \{luiz.wille,jens.kirchner\}@fau.de). J. Kirchner is also with the Department of Information Technology, Fachhochschule Dortmund University of Applied Sciences and Arts, Germany.\\
P. Hofmann, R. Gelfert, and F. Fitzek are with the Deutsche Telekom Chair of Communication Networks, TUD Dresden University of Technology, Germany, and the Centre for Tactile Internet with Human-in-the-Loop (CeTI), Dresden, Germany (e-mail: \{pit.hofmann,randy.gelfert,frank.fitzek\}@tu-dresden.de).\\
Jonas Fuchtmann is with the Minimally Invasive Interdisciplinary Therapeutic Intervention research group, Department of Clinical Medicine, TUM School of Medicine and Health, Technical University Munich, Germany (e-mail: jonas.fuchtmann@tum.de).
}
}
\author{
\IEEEauthorblockN{Alexander Wietfeld\IEEEauthorrefmark{1}, Luiz C.\,P. Wille\IEEEauthorrefmark{1}, Pit Hofmann, Jonas Fuchtmann, Randy Gelfert, Frank H.\,P. Fitzek, \\Wolfgang Kellerer, and Jens Kirchner}
}

\maketitle

\begin{abstract}
Future molecular communication (MC) networks are envisioned to use the human cardiovascular system as a channel. Disease-related changes affect particle transport, allowing an MC receiver to also act as a sensor. In this work, we investigate plaque formation in a carotid bifurcation through a systematic fluid-dynamics simulation study with non-Newtonian blood rheology. We analyze how plaque geometry changes branch-specific particle reception and timing, and use these observations to estimate plaque-induced channel obstruction. At an illustrative reference of only 128 emitted particles, a \qty{250}{\milli\second} observation window and \qty{50}{\milli\second} temporal resolution, a Ridge detector estimates maximum local channel-width reduction with a mean absolute error of \qty{8.30}{\percentagepoint}. We evaluate robustness across system constraints like particle budgets, observation windows, branch access, detection efficiency, count resolution, and synchronization. Coarse counting has little effect when timing features remain available, whereas limited temporal resolution and clock uncertainty substantially affect short-window estimates. Extending the observation window to \qty{1}{\second} reduces these timing-related penalties. The results establish a baseline for plaque sensing from constrained MC observations and identify receiver capabilities relevant to the Internet of Bio-Nano Things.
\end{abstract}


\IEEEpeerreviewmaketitle

\acresetall

\section{Introduction}
Atherosclerosis is characterized by the formation of plaque inside arterial walls, progressively narrowing and obstructing the blood vessel~\cite{Ross1999at}. The carotid artery and its bifurcation are important sites for investigating plaque-related changes~\cite{song2020global}. Monitoring these changes is a potential application of the \ac{IoBNT}, in which bio-nano machines communicate and cooperate to support tasks inside the human body~\cite{Akyildiz2015internet,Kuscu2021internet,Egan2023toward}.

In these \ac{IoBNT} networks, \ac{MC} uses particles transported through blood vessels to convey information between a \ac{TX} and a \ac{RX}. Changes in the propagation environment also affect the received signal. Plaque formation therefore connects \ac{MC} with sensing. By observing changes in particle reception and arrival timing, an \ac{RX} can obtain information about the physical channel through which the particles have traveled.

Felicetti \emph{et al.}~\cite{felicettiModelingCD40BasedMolecular2014a,felicettiSimulationMolecularSignaling2013a} model natural signaling between blood platelets and endothelial cells during plaque formation. Synthetic \ac{MC} offers a complementary perspective by using an engineered communication signal to sense changes in the vessel. Our previous work demonstrates plaque-dependent changes in propagation delay and cumulative particle reception in a simplified straight carotid vessel~\cite{Hofmann2024} as well as the impact of arterial plaque tissue elasticity~\cite{hofmann2025elas}. We subsequently extended this model to non-Newtonian fluid dynamics and pulsatile flow~\cite{Wietfeld2025Plaque}. Wille \emph{et al.}~\cite{Wille2025PulsatileFlow} further investigate the influence of pulsatile flow on nanoparticle propagation in mid-sized arteries.

Sun \emph{et al.}~\cite{Sun2022AtherosclerosisDetection} investigate plaque detection through simulated particle-arrival responses. Shao \emph{et al.}~\cite{Shao2025VascularStenosis} estimate stenosis location and severity from peak-response delay using idealized symmetric, bell-shaped constrictions. Baghel \emph{et al.}~\cite{Baghel2026StenosedArteries} model molecular transport through a symmetric stenosis using one-dimensional drift-diffusion with a spatially averaged flow velocity. Building on these approaches, we study estimation of plaque-induced channel obstruction in a bifurcated channel and explicitly vary the information available to the \ac{RX}.

We investigate which physical plaque properties can be estimated from constrained downstream \ac{MC} observations. Our contributions are (i) characterizing particle routing and arrival timing across 30 plaque geometries in a carotid bifurcation; (ii) comparing compact detectors using particle counts, timing, or both, with complete test geometries excluded from training to capture generalization capabilities; and (iii) evaluating robustness to particle budget, observation duration, and five additional \ac{RX} constraints. This connects plaque-sensitive channel behavior to the information requirements of a sensor.

\section{System Model}\label{sec:system_model}
\begin{figure}[!t]
 \centering
 \includegraphics[width=0.75\linewidth]{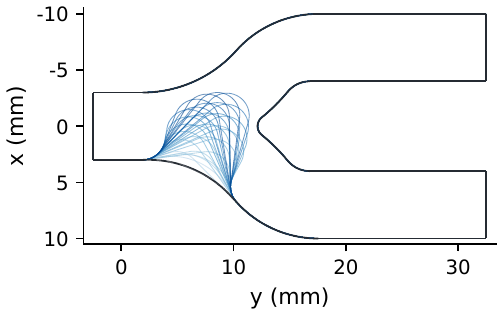}
 \caption{Healthy channel outline and all 30 plaque geometries. Flow runs from left to right. Axes correspond to the simulation coordinates, with $y$ horizontal and $x$ increasing downward.}
 \label{fig:geometry}
\end{figure}

We consider a two-dimensional section of a carotid bifurcation, extending the straight-vessel plaque scenario in~\cite{Hofmann2024,Wietfeld2025Plaque}. The parent vessel has diameter $D_0=\qty{6}{\milli\metre}$ and branches into the \ac{ICA} and \ac{ECA}. Fig.~\ref{fig:geometry} shows the healthy channel as the black outline and the investigated plaque geometries in different shades of blue, ranging from modest plaque sizes to extreme stenosis. In model coordinates, $y$ runs along the parent vessel and $x$ across it. The nominal \ac{CAD} parameters $x_{\text{\acs{CAD}}}$, $y_{\text{\acs{CAD}}}$ and $r_{\text{\acs{CAD}}}$ identify the position and radius parameters used to construct the plaque.

Let $w_0(y)$ and $w_g(y)$ denote the total available channel width in the healthy reference and plaque geometry $g$. We use the maximum local width reduction
\begin{equation}
 S_g=100\max_y\left(1-\frac{w_g(y)}{w_0(y)}\right)
 \label{eq:severity}
\end{equation}
as the primary estimation target, expressed in percent. The supplementary target is the plaque-obstructed channel area
\begin{equation}
 A_g=\int_{\Omega_0\setminus\Omega_g}\mathrm{d}A,
 \label{eq:area}
\end{equation}
where $\Omega_0$ and $\Omega_g$ are the healthy and plaque channel domains.

Particles released upstream propagate to two idealized \acp{RX} at model $y=\qty{32.5}{\milli\metre}$. Negative and positive $x$ identify the \ac{ICA} and \ac{ECA}, respectively. Each particle is assigned its first crossing time $t_i$ relative to the start of injection and its receiving branch. For branch $k\in\{\mathrm{I},\mathrm{E}\}$, the cumulative received fraction  by time $t$ is
\begin{equation}
 C_k(t)=\frac{1}{N_{\mathrm{sim}}}
 \sum_{i=1}^{N_{\mathrm{sim}}}\mathds{1}\{t_i\leq t,\ k_i=k\},
 \label{eq:cumulative}
\end{equation}
where $N_{\mathrm{sim}}$ is the released particle count, $k_i$ is the receiving branch and $\mathds{1}\{\cdot\}$ is the indicator function. Particles without an \ac{RX} crossing have $t_i=\infty$.

\section{Simulation Setup and Dataset}

\subsection{Physiologically-inspired Flow Conditions}
Blood is a non-Newtonian fluid that exhibits a shear-dependent viscosity, i.e., the effective viscosity lowers at higher shear rates $\dot{\gamma}$. Similar to prior work \cite{Hofmann2024, Wietfeld2025Plaque}, the viscosity is calculated based on the Casson fluid model, given as
\begin{equation}
    \mu = \rho \left( \sqrt{\tau_0/\dot{\gamma}} + \sqrt{m} \right)^2
\end{equation}
with constant blood density \cite{Kenner1989measurement} $\rho=\qty{1050}{\kg\per\cubic\metre}$ and coefficients $\tau_0=\qty{2.9e-6}{\m^{2}\s^{-2}}$, $m=\qty{3.93e-6}{\m^{2}\s^{-1}}$.

The average flow in a carotid artery has been measured in \cite{Ackroyd1986CarotidFlow} with a mean of $Q_0=\qty{395}{\milli\litre\per\min}$ for healthy adults. This results in an average blood velocity of 
\begin{equation}
    v_0 = \frac{4Q_0}{\pi D_0^2} = \SI{0.22}{\metre\per\s} \label{eq_carotid_velocity}
\end{equation}
with $D_0=\qty{6}{\milli\metre}$ as the diameter of the carotid as defined in Sec.~\ref{sec:system_model}. In reality, the blood flow will oscillate periodically due to discrete heartbeats, but prior work has shown that often the steady flow rate equivalent can be used to analyze particle spreading over time \cite{Wille2025PulsatileFlow}. Therefore, in this work, we will simplify our simulation approach by only using the steady flow component.

The exact shear rate and viscosity will be part of the simulation results, but we can estimate the wall shear rate based on $\dot{\gamma}=\frac{8 v_0}{D_0} = \qty{284}{\per\s}$, which leads to an average viscosity of $\bar{\mu}=\qty{4.6}{\milli\Pa\s}$.
Based on the given flow parameters, we can estimate the Reynolds number \cite{Hofmann2024} as
$Re = \rho v_0 D_0 / \bar{\mu} = 315$. Since the Reynolds number is well below the critical limit \cite{Hofmann2024} of $\qty{2300}{}$, we assume laminar flow conditions for the simulation scenarios.

\begin{figure}[t]
    \centering
    \includegraphics[width=0.9\linewidth]{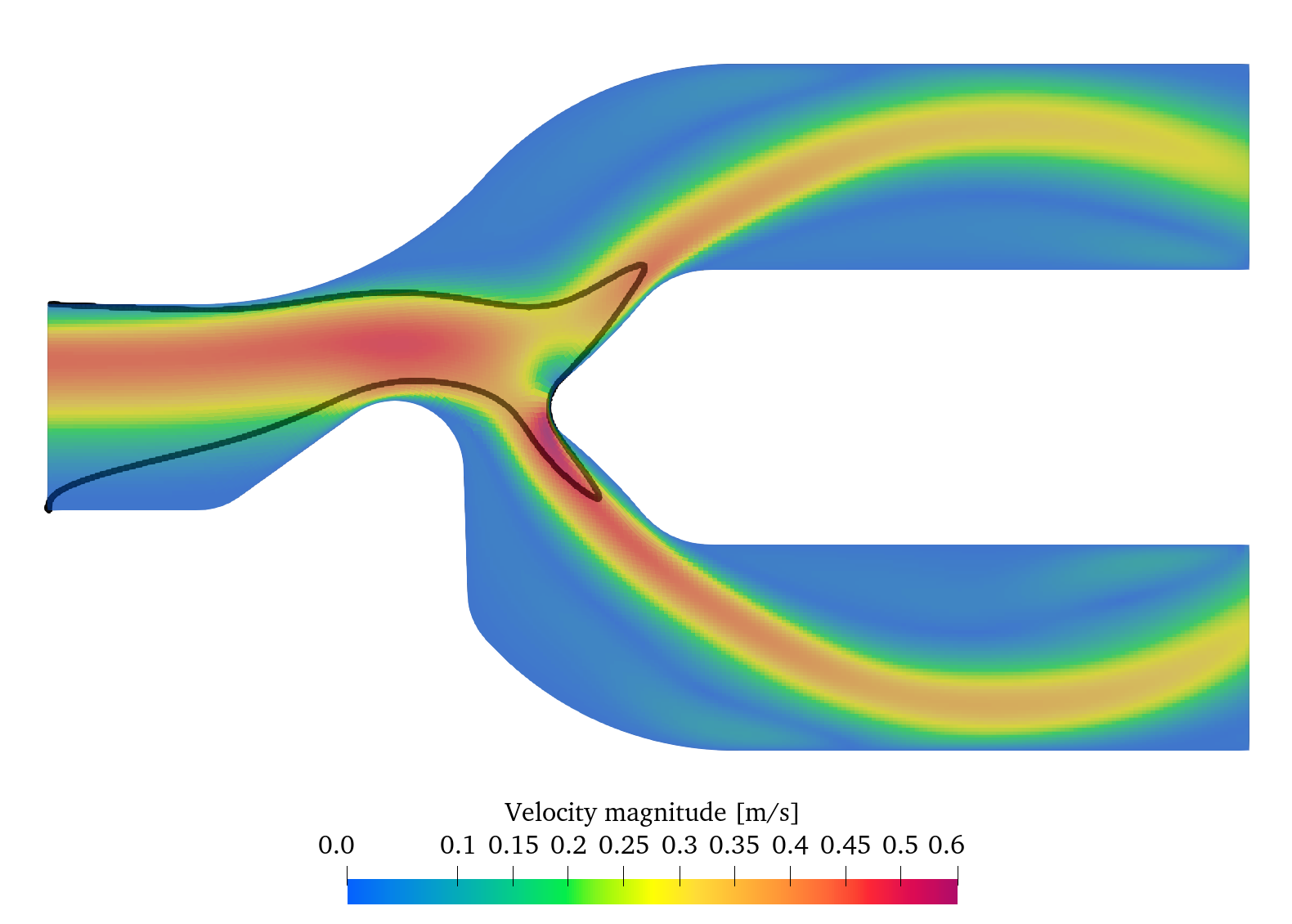}
    \caption{Simulation result of the fluid and particle simulation for time step \SI{0.044}{\s} after particle release. The legend visualizes the fluid velocity magnitude. Black dots represent the particles.}
    \label{fig:flow_sim_particle_exp}
\end{figure}

\subsection{Flow Simulation} \label{sec_flow_sim}
The blood flow in the branching vessel structure is numerically simulated utilizing the \emph{simpleFoam} solver from the OpenFOAM\textsuperscript{\textregistered} environment. This solver uses the Navier-Stokes equations on a discretized mesh to iteratively obtain a balanced flow field. 
In our case, the domain is reduced to a two-dimensional slice through the carotid branch with a depth of \qty{0.1}{\milli\metre}. Thus, the flow is limited to $x-$, and $y-$ direction.

\subsubsection{Boundary Conditions}
The velocity from \eqref{eq_carotid_velocity} is used at the inlet as a constant boundary condition, while it is set to zero at the vessel walls (no-slip boundary condition). For the outlets, a zero-gradient boundary condition (Neumann condition) is used for outgoing flux, while any net backflow (negative flux) is prevented with a zero-value condition. Additionally, we don't impose a fixed velocity profile at the inlet, but let the solver produce an interpolated profile. Due to the curvature features caused by the plaque in our geometry, the velocity profile has the freedom to be skewed; see Fig. \ref{fig:flow_sim_particle_exp}.

For the relative pressure, a reduced Windkessel model~\cite{Xiao2014} is applied at the outlets, and a Neumann boundary condition is set on all other surfaces. The reduced Windkessel model consists of a terminal resistance $R$, which corrects the outlet pressure $p_\text{out}$ based on the solved outlet flow rate $Q_\text{out}$ with
\begin{equation}
    p_\text{out} = R \cdot Q_\text{out} \,.
\end{equation}
The resistance represents the rest of the vessel network beyond our simulated geometry of the carotid branch. This leads to physiologically meaningful pressure and flow rate values. 
Based on values from the literature~\cite{Xiao2014}, we define the terminal resistance of the carotid as $R=\qty{2.15e9}{\Pa\s\per\cubic\metre}$. This value is adjusted with the ratio of the real 3D- to the used 2D- cross section area $R_\text{2D}=(A_\text{3D}/A_\text{2D}) R = \qty{1e11}{\Pa\s\per\cubic\metre}$. Since we have two outlets, the value is applied with factor \qty{2}{} at each outlet (parallel resistance).

\subsection{Particle Simulation}
For the particle simulation, we apply a simple Lagrangian advection scheme~\cite{Xiao2026SPIONS}, where the forces acting on the particles are summed up at each time step to retrieve the particle velocity $v_\text{p}(t)$ and update the particle position $x_\text{p}(t)$.
Similar to other works, the particles are small enough to assume that the fluid drag force is the only major influence on the movement on small time scales~\cite{Wille2025PulsatileFlow}, leading to the following differential equation for the forces acting upon the particles
\begin{equation}
    m_\text{p} \dfrac{\mathrm{d} \vec{v}_\text{p}}{\mathrm{d} t} =  3\pi\mu(\vec{x}_\text{p}) d_\text{p} \left(v_\text{f}(\vec{x}_\text{p}) - \vec{v}_\text{p}(t) \right)
\end{equation}
The particle properties are representing \acp{SPION} \cite{Xiao2026SPIONS} with particle diameter $d_\text{p}=\qty{50}{\nano\metre}$ and single particle density \qty{2047}{\kg\per\cubic\metre}. 
The velocity field $v_\text{f}(\vec{x})$ is obtained through the prior flow simulation (see Sec. \ref{sec_flow_sim}).

The particles are introduced at the carotid inlet at time $t=\qty{0}{s}$ with count $N_\mathrm{sim}=\qty{1e4}{}$ uniformly over the x-coordinate at $y=\qty{-2.5}{\milli\metre}$. The simulation stops after $T_\mathrm{sim}=\qty{5}{\s}$, since most particles will have reached the branch ends by this time.

We employ a very fine boundary layer at the wall with 15 layers decreasing in size towards the wall in order to capture flow gradients very accurately and thus minimizing nonphysical particle wall contacts. If a particle comes into contact with the wall, an elastic rebound is conducted, since we assume non-sticking particle properties. 

\subsection{Dataset and Post-Processing}
The dataset contains 30 plaque geometries and one healthy reference, each with one realized release of $N_{\mathrm{sim}}$ particles. The plaque cases span $S_g=$\qtyrange{10.35}{82.27}{\percent} and $A_g=$\qtyrange{3.93}{39.24}{\square\milli\metre}. Both metrics are calculated from the channel geometry and vary within the investigated family. A dataset with corresponding solutions is openly available under~\cite{PlaqueDataset2026}.

We reconstruct \ac{RX} crossings from saved particle trajectories and express arrival times relative to injection. Full-release channel statistics use the $T_\mathrm{sim}$ analysis window. Smaller particle subsets and shorter observation windows are subsequently used to represent constrained \ac{MC} observations.

\section{Plaque-Dependent Channel Characteristics}
\begin{figure}[t]
 \centering
 \includegraphics[width=0.9\linewidth]{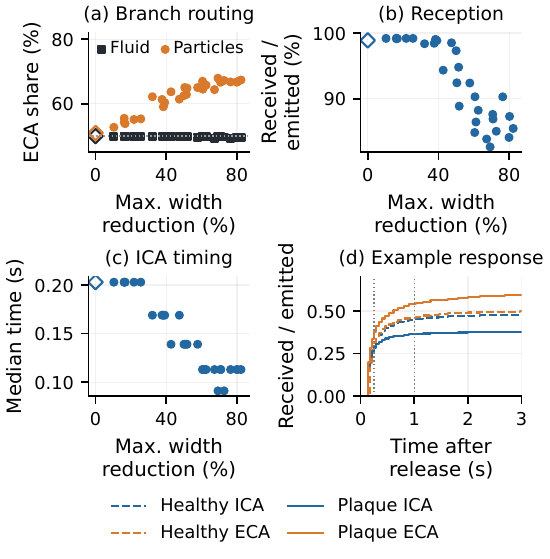}
 \caption{Channel characteristics for full releases of \num{10000} particles. Open diamonds mark the healthy reference. Panels (a)--(c) use the \qty{5}{\second} analysis window. Panel (d) compares the healthy case with a plaque having $S_g=\qty{37.59}{\percent}$ and marks the \qty{0.25}{\second} and \qty{1}{\second} \ac{RX} windows.}
 \label{fig:channel}
\end{figure}

The bifurcation adds branch-specific particle routing to the count and delay changes observed in straight vessels~\cite{Hofmann2024,Wietfeld2025Plaque}. Fig.~\ref{fig:channel}(a) compares the \ac{ECA} share of net outlet fluid flow with its share of received particles. Net fluid flow remains approximately balanced, with an \ac{ECA} share of \qtyrange{49.29}{50.00}{\percent}. Particle routing changes more strongly, with \qtyrange{50.99}{67.94}{\percent} of received particles reaching the \ac{ECA}. Across plaque cases, this fraction increases with the width reduction $S_g$, with a Pearson correlation coefficient of 0.944.

Fig.~\ref{fig:channel}(b) shows that \qtyrange{82.68}{99.22}{\percent} of emitted particles reach either \ac{RX} within \qty{5}{\second}. Reception decreases with increasing width reduction, with correlation $-0.879$. The \ac{ICA} median arrival time also decreases, as shown in Fig.~\ref{fig:channel}(c), with correlation $-0.947$.

Fig.~\ref{fig:channel}(d) illustrates these effects for the healthy channel and a plaque with \qty{37.59}{\percent} maximum width reduction. The plaque response rises earlier and redistributes cumulative reception between the branches. Together, the systematic trends and the example response identify particle routing and arrival timing as complementary sources of information for plaque estimation.

\section{Plaque Detection and Estimation}
\subsection{Receiver Observations and Constraints}
A derived \ac{RX} observation consists of $N$ trajectories sampled without replacement from one simulated release. We generate $J=20$ observations per geometry and setting. These subsets represent particle-sampling variability within the same physical simulation.

The observation horizon $T$ is the deadline after injection, and $\Delta T$ is the \ac{RX} time-bin width. Each particle is detected independently with probability $\eta$. A common clock offset $\delta\sim\mathcal{N}(0,\sigma_{\mathrm{clk}}^2)$ applies to all timestamps in an observation. The reported arrival time is
\begin{equation}
 \widetilde t_i=\Delta T
 \left\lceil\frac{\max(t_i+\delta,0)}{\Delta T}\right\rceil .
 \label{eq:rx_time}
\end{equation}
An arrival contributes when it is detected and both $t_i\leq T$ and $\widetilde t_i\leq T$.

\begin{table}[t]
\vspace*{4pt}
 \caption{Evaluated particle budgets and \ac{RX} constraints.}
 \label{tab:constraints}
 \centering
 \begin{tabular}{lp{0.65\columnwidth}}
 \hline
 Parameter (unit) & Evaluated values\\
 \hline
 $N$ (particles) & 32, 64, 128, 256, 512, 1000, 2500, 5000, 10000\\
 $T$ (\unit{\second}) & 0.25, 0.5, 1, 1.5, 2, 3, 5\\
 $\Delta T$ (\unit{\second}) & 0.025, 0.05, 0.1, 0.25, 0.5, 1\\
 $\eta$ & 1, 0.75, 0.5, 0.25, 0.1\\
 $\sigma_{\mathrm{clk}}$ (\unit{\second}) & 0, 0.025, 0.05, 0.1, 0.25\\
 Count levels $L$ & Exact, 32, 16, 8, 4\\
 Branch access & Both, \ac{ICA} only, \ac{ECA} only, pooled\\
 \hline
 \end{tabular}
\end{table}

Branch access determines whether the \ac{RX} observes both outlets separately, one outlet, or their pooled arrivals without branch identity. Finite count resolution rounds counts to $L$ equally spaced levels between 0 and $N$. Timing features retain the detected timestamps.

Our comparison reference uses $N=128$, $T=\qty{0.25}{\second}$ and $\Delta T=\qty{50}{\milli\second}$, both branches, $\eta=1$, exact counts and $\sigma_{\mathrm{clk}}=0$. We add $T=\qty{1}{\second}$ as a longer-window comparison. Table~\ref{tab:constraints} defines the robustness sweeps around these default conditions.

\subsection{Features, Detectors, and Evaluation}

Let $n_{\mathrm{I}}$ and $n_{\mathrm{E}}$ denote observed branch counts in the \ac{ICA} and \ac{ECA} respectively, and $n=n_{\mathrm{I}}+n_{\mathrm{E}}$. Count features are $n_{\mathrm{I}}$, $n_{\mathrm{E}}$, $n$, the unobserved count $N-n$, received fraction $n/N$, branch fractions $n_{\mathrm{I}}/n$ and $n_{\mathrm{E}}/n$, imbalance $(n_{\mathrm{I}}-n_{\mathrm{E}})/n$, and log ratio $\log[(n_{\mathrm{I}}+0.5)/(n_{\mathrm{E}}+0.5)]$. Timing features comprise each accessible branch's first arrival, 10th, 50th, and 90th percentiles, mean and standard deviation. With both branches accessible, we also include differences of their first arrivals and percentiles. Single-branch and pooled \acp{RX} use only their accessible counts and timing. We compare count-only, timing-only, and combined features.

For each target $\theta_g\in\{S_g,A_g\}$, i.e. reduced channel width or obstructed channel area, the detector maps the available feature vector to an estimate. Ridge is the primary linear baseline, fitted by squared-error minimization with an $\ell_2$ coefficient penalty of weight 1. Robust linear regression uses the Huber loss, which reduces the influence of large residuals, with transition parameter 1.35 and penalty $10^{-3}$. The \ac{KNN} detector averages the targets of the five closest training observations using inverse Euclidean-distance weights. The decision tree uses successive feature thresholds, with maximum depth 3 and at least five observations per leaf. The \ac{MLP} has one hidden layer of 16 units and penalty $10^{-3}$.

All models are fitted separately for each target, feature set, and \ac{RX} condition. Missing features use training-set medians, or zero for an entirely missing column. Features are standardized using training-set means and standard deviations for all models except the tree. Detector settings remain fixed across comparisons.

We partition the 30 plaque geometries into five groups, training on four and testing on the fifth in turn. Every observation of a test geometry is excluded from training. A stricter test excludes all geometries sharing one nominal $x_{\text{\acs{CAD}}}$ level. For predictions $\widehat\theta_{gj}$, we report \ac{MAE} and the coefficient of determination $R^2$,
\begin{align}
 \text{\acs{MAE}}&=\frac{1}{GJ}\sum_{g,j}|\widehat\theta_{gj}-\theta_g|,\\
 R^2&=1-\frac{\sum_{g,j}(\widehat\theta_{gj}-\theta_g)^2}
                    {\sum_{g,j}(\theta_g-\overline\theta)^2},
 \label{eq:metrics}
\end{align}
where $G=30$, $J=20$ and $\overline\theta$ is the mean target across geometries. Width \ac{MAE} is reported in percentage points (\unit{\pp}), and area \ac{MAE} in \unit{\square\milli\metre}. Prediction accuracy, detector and $N,T$ grid plots use one observation seed, while observation-time and \ac{RX}-constraint curves average three observation seeds and three assignments of the five geometry folds.

\subsection{Estimation Results}

\begin{figure}[t]
 \centering
 \includegraphics[width=0.8\linewidth]{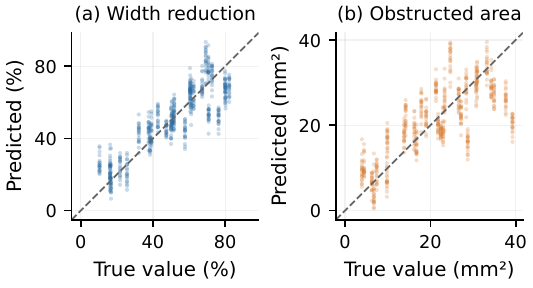}
 \caption{Ridge estimates using counts and timing at the reference setting. Each target includes all 600 test observations from 30 geometries. The dashed line denotes exact prediction.}
 \label{fig:true_vs_pred}
\end{figure}

\begin{figure}[t]
 \centering
 \includegraphics[width=0.7\linewidth]{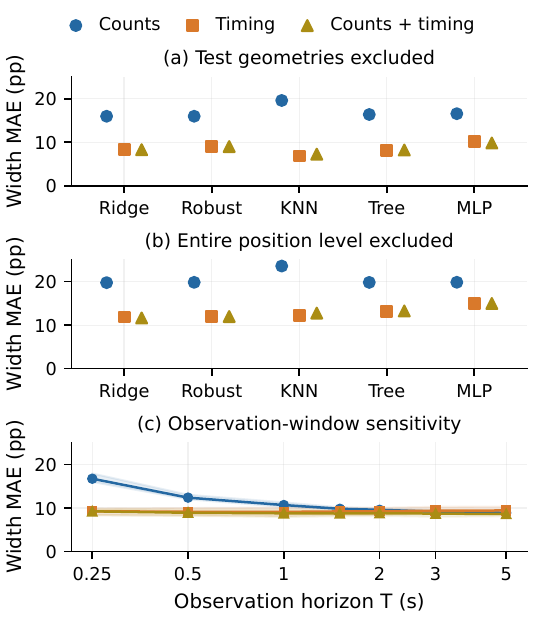}
 \caption{Width-estimation \ac{MAE} for three feature sets. Panels (a,b) compare five detectors at the reference using geometry-separated and omitted-position-level tests. Panel (c) shows Ridge horizon sensitivity, with nine-setting means and min--max bands.}
 \label{fig:detectors}
\end{figure}

Fig.~\ref{fig:true_vs_pred} shows Ridge estimates at the reference setting. Maximum width reduction is estimated with \qty{8.30}{\pp} \ac{MAE} and $R^2=0.749$. For the obstructed area, the corresponding values are \qty{5.26}{\square\milli\metre} and $R^2=0.551$. The true vs. predicted plots show the spread of individual observations around each geometry's true value.

Fig.~\ref{fig:detectors}(a) compares the five detectors. \ac{KNN} achieves \qty{7.27}{\pp} with combined features and \qty{6.85}{\pp} with timing alone. Compact nonlinear processing therefore improves on Ridge for this comparison. Excluding an entire position-parameter level in Fig.~\ref{fig:detectors}(b) increases Ridge's combined-feature error to \qty{11.65}{\pp}. This stricter test highlights the influence of the geometry variation represented in training, but shows that the performance stays relatively stable.

Fig.~\ref{fig:detectors}(c) shows a strong dependence of count-only width estimation on the observation window. Its mean error falls from \qty{16.73}{\pp} at \qty{0.25}{\second} to \qty{10.67}{\pp} at \qty{1}{\second} and \qty{9.58}{\pp} at \qty{2}{\second}. Combined-feature errors are 9.23, 8.77, and \qty{8.80}{\pp}, respectively. Timing provides most of its relative benefit when the \ac{RX} has access only to the early response.

\subsection{Robustness to Receiver Constraints}

\begin{figure}[t]
 \centering
 \includegraphics[width=0.6\linewidth]{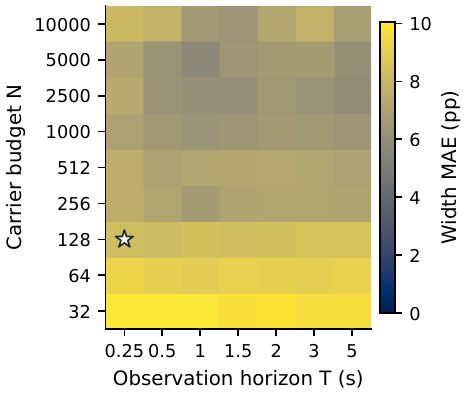}
 \caption{Ridge width \ac{MAE} across all 63 $N,T$ combinations, using combined features and 20 observations per geometry. The star marks the default comparison reference.}
 \label{fig:grid}
\end{figure}

Fig.~\ref{fig:grid} evaluates the information available through particle budget and observation duration. At $T=\qty{0.25}{\second}$, increasing $N$ from 128 to 1000 reduces Ridge width \ac{MAE} from 8.30 to \qty{7.02}{\pp}. The full grid describes sensitivity to these resources, including non-monotonic changes. \ac{MAE} ranges from 5.76 to 10.05 pp across settings, indicating that estimation works across the entire evaluated grid, even for small numbers of particles and short observation times.

\begin{figure*}[h]
 \centering
 \includegraphics[width=0.65\textwidth]{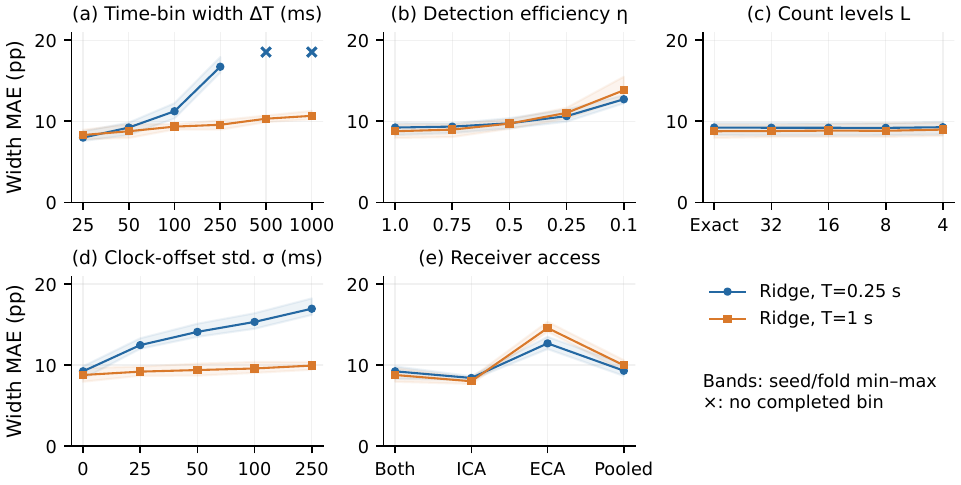}
 \caption{Ridge width \ac{MAE} under five one-factor \ac{RX} constraints at $N=128$. Curves compare $T=\qty{0.25}{\second}$ and \qty{1}{\second}. Bands show seed/fold min--max ranges, not confidence intervals. Crosses mark settings with no completed positive-time bin.}
 \label{fig:constraints}
\end{figure*}

Fig.~\ref{fig:constraints} varies one \ac{RX} factor at a time around a reference point. Temporal resolution and synchronization have the largest short-window effects, i.e., for $T=\qty{0.25}{\second}$. At $\Delta T=\qty{250}{\milli\second}$, Ridge error is \qty{16.73}{\pp} for $T=\qty{0.25}{\second}$ and \qty{9.57}{\pp} for $T=\qty{1}{\second}$. A clock-offset standard deviation of \qty{250}{\milli\second} gives 16.95 and \qty{9.94}{\pp}. Longer observation captures more of the response and reduces the penalties.

At $T=\qty{0.25}{\second}$, four-level counting changes error only from 9.23 to \qty{9.27}{\pp}, whereas detection efficiency $\eta=0.1$ raises it to \qty{12.72}{\pp}. In this scenario, Ridge is robust to coarse counting when timing information remains available. Branch access also matters. Errors are 9.23, 8.41, 12.67 and \qty{9.30}{\pp} for both branches, \ac{ICA} only, \ac{ECA} only and pooled reception, respectively, indicating that the arrival from the branch without the plaque seems to convey more information. Pooled reception also retains performance close to the two-branch reference, making branch identification less critical for this Ridge detector.

Overall, the observation window, accessible branches and timing capabilities jointly determine the estimation performance.

\subsection{Healthy-Reference Plaque Detection}

\begin{figure}[t]
 \centering
 \includegraphics[width=0.7\linewidth]{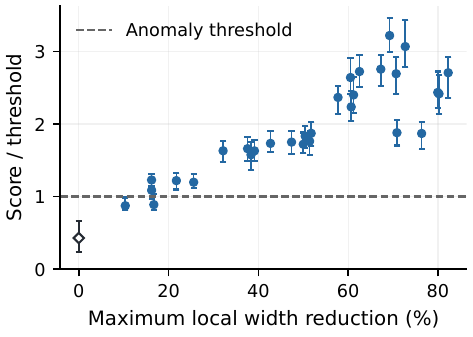}
 \caption{Healthy-reference anomaly score divided by its calibration threshold. Points and whiskers show geometry medians and interquartile ranges across derived observations. The open diamond marks the healthy reference.}
 \label{fig:anomaly}
\end{figure}

We additionally evaluate an exploratory anomaly detector using the healthy release as a reference. Its eight features are received fraction, branch imbalance, and the three arrival-time percentiles per branch divided by $T$. The score is the root-mean-square distance from the mean healthy feature vector.

For each of eight random partitions, disjoint healthy particle pools are used for reference fitting, threshold calibration and evaluation. The threshold is the 95th percentile of calibration scores. Fig.~\ref{fig:anomaly} shows scores normalized by this threshold. At the reference parameter point, \qty{4.17}{\percent} of healthy evaluation observations and \qty{94.13}{\percent} of plaque observations exceed it, with 28 of 30 plaque medians above the threshold, indicating that plaque presence can be differentiated from the healthy case with high likelihood. At $N=1000$ and $T=\qty{5}{\second}$, healthy false positives increase to \qty{15.21}{\percent}, identifying healthy-reference calibration as an important next step.

\subsection{Scope of the Results}
The results describe one parameterized bifurcation under steady flow, with one release per geometry. Width reduction, obstructed area, and cross-channel plaque position vary together. In a separate trial investigation, we find that the evaluated estimators do not reliably recover plaque position $y_\mathrm{CAD}$ along the channel. \ac{RX} time-bin sweeps act on reconstructed trajectories with finite saved-time resolution. Independent releases, broader healthy references, and varied anatomies could extend this baseline and bring it closer to physical validation.

\section{Conclusion and Outlook}

In this paper, we analyzed plaque-dependent particle transport in a carotid bifurcation and estimated channel obstruction from constrained \ac{MC} observations. Branch-specific reception and arrival timing are used as informative features for a set of compact detectors. A robustness analysis identifies temporal resolution and synchronization as important short-window constraints and shows how a longer observation window reduces their influence. Coarse counting can still be effective when timing information remains available. Future work will extend the healthy-reference calibration, evaluate independent releases, varied vascular geometries, and consider experimental receiver implementations.

\printbibliography

@article{Kenner1989measurement,
	doi = {10.1007/bf01907921},
	year = 1989,
	month = Mar,
	publisher = {Springer Science and Business Media {LLC}},
	volume = {84},
	number = {2},
	author = {T. Kenner},
	title = {The Measurement of Blood Density and its Meaning},
	journal = {Basic Res. Cardiol.},
	pages = {111--124}
}

@ARTICLE{Akyildiz2015internet,
  author={Akyildiz, I. F. and Pierobon, M. and Balasubramaniam, S. and Koucheryavy, Y.},
  journal={IEEE Commun. Mag.}, 
  title={{The Internet of Bio-Nano Things}}, 
  year={2015},
  month=Mar,
  volume={53},
  number={3},
  pages={32--40},
  doi={10.1109/MCOM.2015.7060516},}

@article{Kuscu2021internet,
	doi = {10.52953/chbb9821},
	year = 2021,
	month = Dec,
	publisher = {International Telecommunication Union},
	volume = {2},
	number = {3},
	author = {Murat Kuscu and  Bige Deniz Unluturk},
	title = {{Internet of Bio-Nano Things: A} {Review of Applications, Enabling Technologies and Key Challenges}},
	journal = {ITU J. Future Evol. Technol.},
	pages = {1--24}
}

@article{Egan2023toward,
	doi = {10.3390/life13010208},
	year = 2023,
	month = Jan,
	publisher = {{MDPI} {AG}},
	volume = {13},
	number = {1},
	author = {Malcolm Egan and Murat Kuscu and Michael Taynnan Barros and Michael Booth and Antoni Llopis-Lorente and Maurizio Magarini and Daniel P. Martins and Maximilian Schäfer and Pasquale Stano},
	title = {{Toward Interdisciplinary Synergies in Molecular Communications: {Perspectives} from Synthetic Biology, Nanotechnology, Communications Engineering and Philosophy of Science}},
	journal = {Life},
	eid = {208}
}

@article{Ross1999at,
author = {Ross, Russell},
title = {Atherosclerosis — {An} Inflammatory Disease},
journal = {N. Engl. J. Med.},
volume = {340},
number = {2},
year = {1999},
month = Jan,
pages = {115--126},
doi = {10.1056/NEJM199901143400207},
}

@article{song2020global,
  title={Global and Regional Prevalence, Burden, and Risk Factors for Carotid Atherosclerosis: A Systematic Review, Meta-Analysis, and Modelling Study},
  author={Song, Peige and Fang, Zhe and Wang, Hanyu and Cai, Yutong and Rahimi, Kazem and Zhu, Yajie and Fowkes, F. Gerald R. and Fowkes, Freya J. I. and Rudan, Igor},
  journal={Lancet Glob. Health},
  volume={8},
  number={5},
  year={2020},
  month = may,
  pages={e721--e729},
  doi={10.1016/S2214-109X(20)30117-0},
  publisher={Elsevier}
}

@article{felicettiSimulationMolecularSignaling2013a,
  title = {Simulation of Molecular Signaling in Blood Vessels: {{Software}} Design and Application to Atherogenesis},
  shorttitle = {Simulation of Molecular Signaling in Blood Vessels},
  author = {Felicetti, Luca and Femminella, Mauro and Reali, Gianluca},
  year = {2013},
  month = sep,
  journal = {Nano Commun. Netw.},
  volume = {4},
  number = {3},
  pages = {98--119},
  doi = {10.1016/j.nancom.2013.06.002}
}

@article{felicettiModelingCD40BasedMolecular2014a,
  title = {Modeling {{CD40-Based Molecular Communications}} in {{Blood Vessels}}},
  author = {Felicetti, Luca and Femminella, Mauro and Reali, Gianluca and Gresele, Paolo and Malvestiti, Marco and Daigle, John N.},
  year = {2014},
  month = sep,
  journal = {IEEE Trans. NanoBiosci.},
  volume = {13},
  number = {3},
  pages = {230--243},
  doi = {10.1109/TNB.2014.2340134}
}

@article{Hofmann2024, 
title={{A Molecular Communication Perspective on Detecting Arterial Plaque Formation}}, 
volume={10}, 
DOI={10.1109/tmbmc.2024.3423005}, 
number={3}, 
journal={IEEE Trans. Mol. Biol. Multi-Scale Commun.}, 
author={Hofmann, Pit and Schmidt, Sebastian and Wietfeld, Alexander and Zhou, Pengjie and Fuchtmann, Jonas and Fitzek, Frank H. P. and Kellerer, Wolfgang}, 
year={2024}, 
month=sep, 
pages={458--463} }

@inproceedings{Wietfeld2025Plaque,
  title = {Advanced {{Plaque Modeling}} for {{Atherosclerosis Detection Using Molecular Communication}}},
  booktitle = {Proc. IEEE ICC},
  author = {Wietfeld, Alexander and Hofmann, Pit and Fuchtmann, Jonas and Zhou, Pengjie and Zheng, Ruifeng and Cabrera, Juan A. and Fitzek, Frank H.P. and Kellerer, Wolfgang},
  year = 2025,
  month = jun,
  pages = {6056--6062},
  doi = {10.1109/ICC52391.2025.11161338},
}

@article{Ackroyd1986CarotidFlow,
  title = {Quantitative Common Carotid Artery Blood Flow: {{Prediction}} of Internal Carotid Artery Stenosis},
  shorttitle = {Quantitative Common Carotid Artery Blood Flow},
  author = {Ackroyd, Nigel and Gill, Robert and Griffiths, Kaye and Kossoff, George and Appleberg, Michael},
  year = 1986,
  month = jun,
  journal = {J. Vasc. Surg.},
  volume = {3},
  number = {6},
  pages = {846--853},
  issn = {07415214},
  doi = {10.1016/0741-5214(86)90148-5}
}

@article{Wille2025PulsatileFlow,
  title = {From {{Steady}} to {{Pulsatile Flow}} in {{Molecular Communication}}: {{Propagation}} of {{Nanoparticles}} in {{Mid-Sized Arteries}}},
  shorttitle = {From {{Steady}} to {{Pulsatile Flow}} in {{Molecular Communication}}},
  author = {Wille, Luiz C. P. and Pfannenm{\"u}ller, Christof and Kirchner, Jens},
  year = 2025,
  month = dec,
  journal = {IEEE Trans. Mol. Biol. Multi-Scale Commun.},
  volume = {11},
  number = {4},
  pages = {531--536},
  issn = {2372-2061, 2332-7804},
  doi = {10.1109/TMBMC.2025.3608558}
}

@article{Sun2022AtherosclerosisDetection,
  author  = {Sun, Yue and Liu, Meiling and Xiao, Yue and Chen, Yifan},
  title   = {A Novel Molecular Communication Inspired Detection Method for the Evolution of Atherosclerosis},
  journal = {Comput. Methods Programs Biomed.},
  volume  = {219},
  eid     = {106756},
  year    = {2022},
  month=jun,
  doi     = {10.1016/j.cmpb.2022.106756}
}

@article{Shao2025VascularStenosis,
  author  = {Shao, Zitong and Zhang, Pengfei and Wang, Xiaofang and Lu, Pengfei},
  title   = {The Modeling and Detection of Vascular Stenosis Based on Molecular Communication in the Internet of Things},
  journal = {J. Sens. Actuator Netw.},
  volume  = {14},
  number  = {5},
  eid     = {101},
  year    = {2025},
  month=oct,
  doi     = {10.3390/jsan14050101}
}

@article{Baghel2026StenosedArteries,
  author  = {Baghel, Aakanksha and Das, Debanjan and Mahapatra, Rajarshi},
  title   = {Communication-Inspired Modeling of Molecular Transport Dynamics in Stenosed Arteries Toward {IoBNT} System},
  journal = {IEEE Commun. Lett.},
  volume  = {30},
  pages   = {2048--2052},
  year    = {2026},
  month=may,
  doi     = {10.1109/LCOMM.2026.3695150}
}

@article{Xiao2026SPIONS,
  title = {Physics of {{SPION}}--{{Surface Interactions}} in {{Fluidic Channels}}: {{Implications}} for {{Molecular Communication Systems}}},
  shorttitle = {Physics of {{SPION}}--{{Surface Interactions}} in {{Fluidic Channels}}},
  author = {Xiao, Keyu and Thalmayer, Angelika S. and Kirchner, Jens and Fischer, Georg},
  year = 2026,
  journal = {IEEE Trans. Mol. Biol. Multi-Scale Commun.},
  volume = {12},
  pages = {504--509},
  month=apr,
  issn = {2372-2061, 2332-7804},
  doi = {10.1109/TMBMC.2026.3686303}
}

@article{Xiao2014,
  title = {A Systematic Comparison between 1-{{D}} and 3-{{D}} Hemodynamics in Compliant Arterial Models},
  author = {Xiao, Nan and Alastruey, Jordi and Figueroa, C. Alberto},
  year = 2014,
  month = feb,
  journal = {Int. J. Numer. Methods Biomed. Eng.},
  volume = {30},
  number = {2},
  pages = {204--231},
  issn = {2040-7939, 2040-7947},
  doi = {10.1002/cnm.2598},
}

@inproceedings{hofmann2025elas,
  title = {{The Impact of Arterial Plaque Tissue Elasticity on
In-Body Molecular Communication Scenarios}},
  booktitle = {Proc. 9th Workshop on Molecular Communications},
  author = {Pit Hofmann and Alexander Wietfeld and Jonas Fuchtmann and Pengjie Zhou and Ruifeng Zheng and Juan A. Cabrera and Frank H.P. Fitzek and Wolfgang Kellerer},
  year = 2025,
  month = apr,
  url = {https://molecularcommunications.org/wp-content/uploads/2025/05/3-5_Hofmann.pdf},
}

@misc{PlaqueDataset2026,
  author = {Alexander Wietfeld and Luiz C. P. Wille and Pit Hofmann and Jonas Fuchtmann and Frank H. P. Fitzek and Jens Kirchner and Wolfgang Kellerer},
  title = {Dataset - Molecular Communication for Plaque Sensing in Vascular Bifurcations: Channel Effects and Receiver Constraints},
  year = {2026},
  url = {https://doi.org/10.5281/zenodo.23082665},
  type = {Dataset}
}
\end{document}